\documentclass[conference,a4paper]{IEEEtran}

\usepackage{graphicx}
\usepackage{enumerate}
\usepackage[cmex10]{amsmath}
\usepackage{cite}
\usepackage{bm}
\usepackage[dvipsnames]{xcolor}
\usepackage{amsmath,amssymb,amsthm}
\usepackage{stfloats}
\usepackage{suffix}
\usepackage{mathtools}
\usepackage{dsfont}
\usepackage[hidelinks]{hyperref}
\usepackage{balance} 

\def\bA{{\mathbf{A}}}  \def\bC{{\mathbf{C}}}  \def\bE{{\mathbf{E}}}
\def\bf{{\mathbf{F}}} \def\bG{{\mathbf{G}}} \def\bH{{\mathbf{H}}} \def\bI{{\mathbf{I}}} \def\bp{{\mathbf{J}}}
    
 \def\bq{{\mathbf{Q}}}   
\def\bU{{\mathbf{U}}} \def\bV{{\mathbf{V}}}   

\def\ba{{\mathbf{a}}}  \def\bc{{\mathbf{c}}}  
\def\bf{{\mathbf{f}}}

\def\bp{{\mathbf{p}}} \def\bq{{\mathbf{q}}}  \def\bs{{\mathbf{s}}} 
    \def\by{{\mathbf{y}}}
\def\bz{{\mathbf{z}}}

\makeatletter
\def\blfootnote{\xdef\@thefnmark{}\@footnotetext}
\makeatother

\begin{document}
\title{On the Sum Secrecy Rate of Multi-User Holographic MIMO Networks}
% \title{Multi-User Holographic MIMO Networks: Enhancing the Secrecy Capacity}
\author{Arthur S. de Sena$^{1,\ddag}$, Jiguang~He$^2$, Ahmed Al Hammadi$^2$, Chongwen Huang$^3$, \\Faouzi Bader$^2$, Merouane Debbah$^4$, Mathias Fink$^5$\\
        $^1$Centre for Wireless Communications, FI-90014, University of Oulu, Finland\\
        $^2$Technology Innovation Institute, 9639 Masdar City, Abu Dhabi, UAE\\ 
        $^3$College of Information Science and Electronic
Engineering, Zhejiang University, Hangzhou 310027, China\\
        $^4$Khalifa University of Science and Technology, P O Box 127788, Abu Dhabi, UAE\\
        $^5$Institut Langevin, ESPCI Paris, Université PSL, CNRS, 75005, Paris, France}
 \maketitle

\begin{abstract}
The emerging concept of extremely-large holographic multiple-input multiple-output (HMIMO), beneficial from compactly and densely packed cost-efficient radiating meta-atoms, has been demonstrated for enhanced degrees of freedom even in pure line-of-sight conditions, enabling tremendous multiplexing gain for the next-generation communication systems. Most of the reported works focus on energy and spectrum efficiency, path loss analyses, and channel modeling. The extension to secure communications remains unexplored. In this paper, we theoretically characterize the secrecy capacity of the HMIMO network with multiple legitimate users and one eavesdropper while taking into consideration artificial noise and max-min fairness. We formulate the power allocation (PA) problem and address it by following successive convex approximation and Taylor expansion. We further study the effect of fixed PA coefficients, imperfect channel state information, inter-element spacing, and the number of Eve's antennas on the sum secrecy rate. Simulation results show that significant performance gain with more than 100\% increment in the high signal-to-noise ratio (SNR) regime for the two-user case is obtained by exploiting adaptive/flexible PA compared to the case with fixed PA coefficients. 
\end{abstract}
%and the number of Eve's antennas
\blfootnote{\ddag During the time of this research, A. S. Sena was affiliated with the Technology Innovation Institute, 9639 Masdar City, Abu Dhabi, UAE. He is now with the Centre for Wireless Communications, FI-90014, University of Oulu, Finland.}

% \footnote{\ddag \ During the time of this research, the lead author was affiliated with the Technology Innovation Institute, 9639 Masdar City, Abu Dhabi, UAE. His current affiliation is with the Centre for Wireless Communications, FI-90014, University of Oulu, Finland.}

\section{Introduction}
Secure transmissions have always been desired in wireless communications. However, due to the broadcast nature of the wireless propagation, challenges arise in secured transmissions. In the literature, researchers focused on physical layer security from the information-theoretic perspective and introduced artificial noise (AN) to guarantee that all the legitimate users have a higher rate than the eavesdroppers, complementary to traditional complex cryptographic approaches~\cite{Zhou2009, Oggier2011}. Under the framework of multiple-input multiple-output (MIMO), the design of AN is usually jointly considered with precoder design and resource allocation, such as transmit power allocation (PA). By extending from the reconfigurable intelligent surface (RIS) free MIMO network to the RIS assisted one, RIS was verified to bring more flexibility and obvious performance enhancement~\cite{Hong2020,Rafieifar2023}. However, in RIS assisted MIMO networks, due to its passive property, channel state information acquisition becomes, if not infeasible, inevitably difficult, which in turn harms secrecy performance.    

Recently, the active counterpart of RIS, termed as holographic MIMO (HMIMO), serves as a transceiver with a low-cost transformative wireless planar structure comprising of densely packed sub-wavelength metallic or dielectric scattering particles, which is capable of shaping electromagnetic waves according to specific requirements. It is a promising candidate technology for 6G, offering a cost-effective and energy-efficient way of realizing the extremely large-scale MIMO (XL-MIMO)~\cite{Huang2020}. With the introduction of sub-wavelength inter-element spacing, many good properties can be found, e.g., large degrees of freedom (DoFs) even under the condition of line of sight (LoS) connectivity. HMIMO is in favor of near-field communications, standing out in millimeter wave (mmWave) and Terahertz (THz) communications with vast available bandwidths but short communication range. Regarding HMIMO, many reported works focused on channel modeling, beamforming design, and resource allocation~\cite{Pizzo21,Ji23,Pizzo20,Demir2022,Wei2022}. However, secure communication in HMIMO network has not yet been investigated. 

In this paper, we analyze the secrecy performance of the multi-user HMIMO networks with the introduction of AN. In the analysis, we simplify the process by decoupling the following three tasks: (i) the design of base station (BS) transmit beamforming, (ii) that of receive filter, and (iii) PA between the information symbol and AN, and accordingly propose a multi-stage approach for secrecy analysis under the assumption of imperfect channel state information (CSI).  For the PA, we follow the max-min fairness (MMF), which has already been applied in networking level power control in massive MIMO~\cite{Yang2017}, and formulate the optimization problem, which aims at finding the optimal PA between the desired information signals and AN as to reach the maximal sum secrecy rate. We examine the effect of various system parameters, e.g., imperfectness of the CSI and inter-element spacing, on the sum secrecy rate of the studied system. The proposed PA approach is verified to outperform the case with fixed PA coefficients.

\textit{Notations}: Bold lowercase letters denote vectors (e.g., $\ba$), while bold capital letters represent matrices (e.g., $\bA$). The operators $(\cdot)^\mathsf{T}$ and $(\cdot)^\mathsf{H}$ denote transpose and Hermitian transpose, respectively. $\mathrm{diag}(\ba)$ denotes a square diagonal matrix with the entries of $\ba$ on its main diagonal, $\mathbf{0}$ denotes the all-zero vector or matrix, $\bI_{M}$ ($M\geq2$) denotes the $M\times M$ identity matrix, and $j = \sqrt{-1}$. $\|\cdot\|_2$ denotes the Euclidean norm of a vector, and $|\cdot|$ returns the absolute value of a complex number. $[\ba]_m$ and $[\bA]_{:, m}$ denote the $m$-th element of $\ba$ and $m$-th column of $\bA$. 
% Finally, $\Re\{\cdot\}$ returns the real part of its complex argument. 

% \subsection{Related Work}

% \subsection{Motivations and Contributions}

\section{System Model}\label{sec_Sys_Model}
\begin{figure}[t]
	\centering
\includegraphics[width=.7\linewidth]{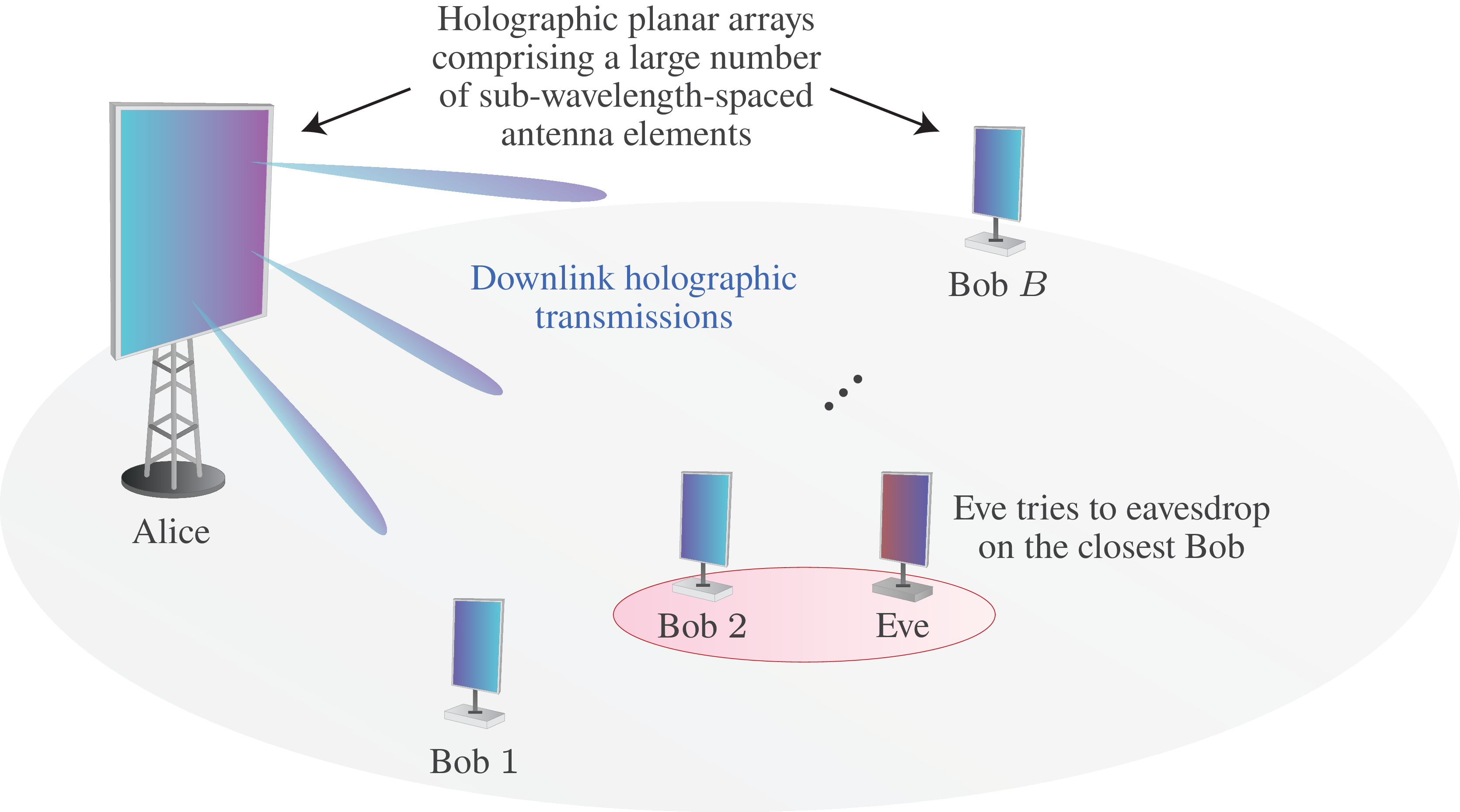}
	\caption{Multi-user HMIMO transmission system with an eavesdropper in the proximity of the legitimate users.}
		\label{System_model}
\end{figure}

We consider a downlink transmission scenario where one BS (a.k.a. Alice) communicates with multiple legitimate users (a.k.a. Bobs) concurrently in the presence of one eavesdropper (a.k.a. Eve). Specifically, we assume the existence of $B$ Bobs in the system, indexed by the set $\mathcal{B} = \{1, \cdots, B \}$, and that all communication nodes are equipped with a holographic uniform planar array (UPA), as illustrated in Fig. \ref{System_model}. The antenna arrays of Alice and Eve comprise $N_{\text{A}} = N_{\text{A}, x} \times N_{\text{A},y}$ and $N_{\text{E}} = N_{\text{E},x} \times N_{\text{E},y}$ antenna elements, respectively, and without loss of generality all the $B$ Bobs are equipped with an equal number of antennas $N_{\text{B}} = N_{\text{B}, x} \times N_{\text{B},y}$, i.e., $ N_b = N_{\text{B}}, \forall b \in \mathcal{B}$, 
in which $\{N_{\text{A}, x}, N_{\text{B}, x}, N_{\text{E}, x}\}$ and $\{N_{\text{A}, y}, N_{\text{B}, y}, N_{\text{E}, y}\}$ correspond to the number of elements in the $x$-axis and $y$-axis directions, respectively. Moreover, the inter-element spacing in all antenna arrays, denoted by $\delta$, is set to less than half wavelength $\lambda$, i.e., $\delta < \lambda/2$. As a result, the lengths of the arrays in the $x$-axis and $y$-axis directions for the $i$th communication node are given by $L_{i,x} = N_{i,x} \delta$ and $L_{i,y} = N_{i,y} \delta$, where the coordinates in $\mathbb{R}^3$ of all antenna elements are organized into the matrix $\mathbf{C}_i = [\mathbf{c}_{i,1}, \cdots, \mathbf{c}_{i, N_i}]\in \mathbb{R}^{3\times N_i}$, and the vector $\mathbf{c}_{i,n} \in \mathbb{R}^3$ corresponds to the three dimensional (3D) position of the $n$-th antenna element, for $n = 1, \cdots, N_i$, with $i\in \{\text{A}, \mathcal{B}, \text{E}\}$.

\subsection{Channel Model}

We employ the electromagnetic compliant channel model for HMIMO communications from~\cite{Pizzo21}, which accurately approximates the HMIMO electromagnetic multi-path propagation through an asymptotic Fourier transform-based Karhunen-Loeve channel expansion. More specifically, the wireless channel between Alice and the $u$-th user, for $u \in \{\mathcal{B}, \text{E}\}$, i.e., valid for all the Bobs and Eve, can be given by
\begin{equation}\label{ch_eq_01}
    \bH_{u} = \bm{\Phi}_{u}\tilde{\bH}_{u}\bm{\Phi}_{\text{A}}^\mathsf{H} \in \mathbb{C}^{N_u \times N_{\text{A}}},
\end{equation}
where $\bm{\Phi}_{u} \in \mathbb{C}^{N_u \times n_u}$ and $\bm{\Phi}_{\text{A}} \in \mathbb{C}^{N_{\text{A}} \times n_{\text{A}}}$ are semi-unitary matrices, i.e.,  $\bm{\Phi}^\mathsf{H}_{u}\bm{\Phi}_{u} = \bI_{n_u}$ and $\bm{\Phi}^\mathsf{H}_{\text{A}}\bm{\Phi}_{\text{A}} = \bI_{n_{\text{A}}}$, comprising the array response vectors $\bm{\theta}(l_{u,x}, l_{u,y}, \bC_{u}) \in \mathbb{C}^{N_u}$ and $\bm{\theta}(l_{\text{A},x}, l_{\text{A},y},  \bC_\text{A}) \in \mathbb{C}^{N_{\text{A}}}$ of the $u$-th user and Alice, respectively, in which the $n$-th entry of $\bm{\theta}(l_{i,x}, l_{i,y}, \bC_{i})$, for $n = 1, \cdots, N_i$, and $i \in \{\text{A}, \mathcal{B}, \text{E}\}$, can be computed by
\begin{align}
    [\bm{\theta}(l_{i,x}, l_{i,y}, \bC_{i})]_n &= \frac{1}{\sqrt{N_i}} e^{j \left(\left[\frac{2\pi}{L_{i,x}}l_{i,x}, \frac{2\pi}{L_{i,y}}l_{i,y}, \gamma(l_{i,x},l_{i,y}) \right] \bc_{i,n} \right)}, \nonumber
\end{align}
where $\gamma(l_{i,x},l_{i,y}) = \sqrt{\kappa^2 - \left(\frac{2\pi}{L_{i,x}}l_{i,x} \right)^2 - \left(\frac{2\pi}{L_{i,y}}l_{i,y} \right)^2 }$ with $\kappa = \frac{2\pi}{\lambda}$ denoting the wavenumber of the system, and $l_{i,x}$ and $l_{i,y}$ are the sampling points in the wavenumber domain, which lead to non-zero angular responses only when the points are within the lattice ellipse \cite{Pizzo21} $\mathcal{E}_{i} = \left\{(l_{i,x},l_{i,y}) \in \mathbb{Z}^2 : \left( \frac{\lambda}{L_{i,x}} l_{i,x} \right)^2 + \left( \frac{\lambda}{L_{i,y}} l_{i,y} \right)^2 \leq 1\right\}$. In particular, with a uniform sampling, these points can be obtained through $l_{i,x} \in \mathcal{E}_{i,x} = \left\{ \left\lceil \frac{-L_{i,x} + (q_{i,x}-1) \lambda}{\lambda} \right\rceil \right\}$, for $q_{i,x} = 1, \cdots, \left\lceil \frac{L_{i,x}}{\lambda}\right\rceil$, and $l_{i,y} \in \mathcal{E}_{i,y} = \left\{ \left\lceil \frac{-L_{i,y} + (q_{i,y}-1) \lambda}{\lambda} \right\rceil \right\}$, for $q_{i,y} = 1, \cdots, \left\lceil \frac{L_{i,y}}{\lambda}\right\rceil$, which results in $n_i = 4 \left\lceil \frac{L_{i,x}L_{i,y}}{\lambda^2}\right\rceil$, for $u \in \{\text{A}, \mathcal{B}, \text{E}\}$ \cite{Ji23}.
% , where $n_i$ provides a theoretical upper bound for the number of degrees of freedom (DoFs) of the HMIMO channel $\bH_{u}, u \in \{\mathcal{B}, \text{E}\}$ \cite{Ji23}. 
Moreover, the matrix $\tilde{\bH}_{u}$ collects the small-scale fading coefficients in the angular domain, which can be structured as $\tilde{\bH}_{u} = \bm{\Sigma}_{u} \odot \bG_{u} \in \mathbb{C}^{n_u \times n_{\text{A}}}$, where $\bG_{u} \in \mathbb{C}^{n_u \times n_{\text{A}}}$ is a random matrix with entries following the complex Gaussian distribution with zero mean and unity variance, and $\bm{\Sigma}_{u} \in \mathbb{R}^{n_u \times n_{\text{A}}}$ is
a matrix that collects $n_u \times n_{\text{A}}$ scaled standard deviations $\{\sqrt{N_{\text{A}} N_u}\sigma(l_{u,x}, l_{u,y}, l_{\text{A},x}, l_{\text{A},y})\}$ of the channel, where the variances $\sigma^2(\cdot)$'s describe the power transferred from Alice to the $u$-th receiver in the corresponding wavenumber sampling points, with $u \in \{\mathcal{B}, \text{E}\}$. Under the assumption of isotropic scattering, the variances observed at Alice and receivers can be decoupled, i.e., $\sigma^2(l_{u,x}, l_{u,y}, l_{\text{A},x}, l_{\text{A},y}) \approx \sigma^2(l_{u,x}, l_{u,y}) \sigma^2(l_{\text{A},x}, l_{\text{A},y})$. Thus, $\sigma^2(l_{i,x}, l_{i,y})$, for $i \in \{\text{A}, \mathcal{B}, \text{E}\}$, can be calculated as follows \cite{Pizzo20}
\begin{equation}
    \sigma^2(l_{i,x}, l_{i,y}) \hspace{-0.1cm} = \hspace{-0.1cm}\frac{1}{4\pi}\hspace{-0.1cm} \int_{\frac{\lambda}{L_{i,x}}l_{i,x}}^{\frac{\lambda}{L_{i,x}}(l_{i,x}+1)} \hspace{-0.3cm}\int_{\frac{\lambda} {L_{i,y}}l_{i,y}}^{\frac{\lambda}{L_{i,y}}(l_{i,y}+1)}  \hspace{-0.4cm}   \frac{\mathds{1}_{\mathcal{D}} (x,y)}{\sqrt{1 - x^2 - y^2}} dx dy,  
\end{equation} 
for $l_{i,x} \in \mathcal{E}_{i,x}$ and $l_{i,y} \in \mathcal{E}_{i,y}$, where $\mathcal{D} = \{(x,y) \in \mathbb{R}^2 : x^2 + y^2 \leq 1\}$ is a disk of radius $1$ centered at the origin. The closed-form expression of $\sigma^2(l_{i,x}, l_{i,y})$ is derived in \cite[Appendix IV.C]{Pizzo20}. As a result, the matrix of standard deviations can be obtained as
\begin{equation}
    \bm{\Sigma}_u = \bm{\sigma}_u \bm{\sigma}_{\text{A}}^\mathsf{T}, \qquad \text{ for }\; u \in \{\mathcal{B}, \text{E}\},
\end{equation}
where $\bm{\sigma}_i \in \mathbb{R}^{n_i}$ is the vector that collects the standard deviations $\{\sqrt{N_i}\sigma(l_{i,x}, l_{i,y})\}, \forall l_{i,x} \in \mathcal{E}_{i,x}, \forall l_{i,y} \in \mathcal{E}_{i,y}$, for $i \in \{\text{A}, \mathcal{B}, \text{E}\}$.

Recall that $\bm{\Phi}_{u}$ and $\bm{\Phi}_{\text{A}}$ are deterministic, depending only on the structure of the antenna arrays. Also, the entries of $\bm{\Sigma}_{u}$ change slowly compared to the coherence interval of the fast-fading channel coefficients. Given these facts, we assume that $\bm{\Phi}_{u}$, $\bm{\Phi}_{A}$, and $\bm{\Sigma}_{u}$ are perfectly known in the system. However, we introduce imperfectness on $\bG_u$, which is modeled by a first-order Gauss-Markov process
\begin{align}\label{imp_csi}
    \hat{\bG}_u = \sqrt{1-\xi^2}\bG_u + \xi \bE_u, \vspace{-0.25cm}
\end{align}
where $\bE_u$ is a complex standard Gaussian distributed error matrix, and $\xi^2$ represents the variance of the channel estimation error. The effect of  imperfect $\bG_u$ on secrecy performance will be evaluated comprehensively in Section~\ref{sec_Sim_results}.

\subsection{Signal Model}

Under the above channel model, Alice transmits an information symbol $s_{b}$ to the $b$-th Bob, $\forall b \in \mathcal{B}$. We assume that Alice does not have any knowledge of the channel or location information of Eve. As a result, it becomes challenging to avoid information leakage to Eve through beamforming only. To mitigate this security threat, Alice superimposes a random AN $w_b \in \mathbb{C}$ onto the information symbol of each Bob, satisfying $\mathrm{E}\{|w_b|^2\}= 1$. More specifically, Alice transmits the following beamformed data stream
\begin{equation}
\bs = \sum_{b = 1}^{B} \bf_b (\sqrt{\alpha_{b}} s_b + \sqrt{\beta_{b}} w_b) \in \mathbb{C}^{N_{\text{A}}},\vspace{-0.25cm}
\end{equation}
where $\bf_b \in \mathbb{C}^{N_{\text{A}}}$ is the beamforming vector for the $b$-th Bob, such that $\|\bf_b\|_2^2 = 1$, $\alpha_{b}$ and $\beta_{b}$ are the PA coefficients for the information symbol and AN, respectively, with a total transmit power constraint $P_T = \sum_{b=1}^B \alpha_{b} + \beta_{b}$. Furthermore, the information symbols $s_{b}$'s are assumed to have zero mean and unity variance, i.e., $\mathrm{E}\{ |s_{b}|^2\} = 1$. With these assumptions, the signals received by the $b$-th user and Eve can be written, respectively, as
\begin{align}
    \by_b &= \bH_b\sqrt{\zeta_b}\sum_{k\in \mathcal{B}} \bf_k (\sqrt{\alpha_k} s_k + \sqrt{\beta_k} w_k) + \bz_b \in \mathbb{C}^{N_{\text{B}}},\\
    \by_E &= \bH_{\text{E}}\sqrt{\zeta_{\text{E}}}\sum_{k\in \mathcal{B}}\bf_k (\sqrt{\alpha_k} s_k + \sqrt{\beta_k} w_k) + \bz_{\text{E}} \in \mathbb{C}^{N_{\text{E}}},\vspace{-0.35cm}
\end{align}
where $\zeta_b = d_b^{-\eta}\Lambda$ and $\zeta_{\text{E}} = d_{\text{E}}^{-\eta}\Lambda$ model the large-scale fading coefficients, in which $d_b$ and $d_{\text{E}}$ denote the distances from Alice to the $b$-th Bob and Eve, respectively, $\eta$ represents the path-loss exponent, and $\Lambda$ is the array gain parameter. Moreover, $\bz_b$ and $\bz_{\text{E}}$ are the corresponding additive noise vectors, whose entries follow the complex Gaussian distribution with zero mean and variance $\sigma^2_z$.

\subsection{Transmit Beamformer Design}
In this subsection, we focus on the design of $\bf_b \in \mathbb{C}^{N_{\text{A}}}$, $\forall b \in \mathcal{B}$. Specifically, we wish to avoid information leakage to non-intended Bobs. Before introducing the beamforming design, we expand the HMIMO channel model in Eq.~\eqref{ch_eq_01} as follows 
\begin{align}\label{ch_eq_02}
    \bH_b &= \bm{\Phi}_{b} \left( \bm{\Sigma}_{b} \odot \bG_{b} \right) \bm{\Phi}_{\text{A}}^\mathsf{H} = \bm{\Phi}_{b} \left( \left[\bm{\sigma}_b \bm{\sigma}_{\text{A}}^\mathsf{T} \right] \odot \bG_{b} \right) \bm{\Phi}_{\text{A}}^\mathsf{H} 
    \nonumber\\
    &=
    \bm{\Phi}_{b} \mathrm{diag}(\bm{\sigma}_b)  \bG_{b}  \mathrm{diag}(\bm{\sigma}_{\text{A}}) \bm{\Phi}_{\text{A}}^\mathsf{H} = \bm{\Phi}_{b} \bm{\Delta}_b  \bG_{b}  \bm{\Delta}_{\text{A}} \bm{\Phi}_{\text{A}}^\mathsf{H},
\end{align}
where $\bm{\Delta}_b \triangleq  \mathrm{diag}(\bm{\sigma}_b)$ and $\bm{\Delta}_{\text{A}} \triangleq  \mathrm{diag}(\bm{\sigma}_{\text{A}})$. Given the expansion in Eq.~\eqref{ch_eq_02} and the aforementioned property  $\bm{\Phi}_{\text{A}}^\mathsf{H}\bm{\Phi}_{\text{A}} = \bI_{n_{\text{A}}}$, we can design the desired beamforming vector with the following structure $\bf_b = \bm{\Phi}_{\text{A}} \bp_b$, where $\bp_b \in \mathbb{C}^{n_{\text{A}}}$ is an inner beamformer computed based on the null space spanned by the reduced-dimension effective matrices of unintended users given by $\bm{\Phi}_{\text{A}}^\mathsf{H}\bH_{b'}^\mathsf{H} =   
\bm{\Delta}_{\text{A}}^\mathsf{H} \bG_{b'}^\mathsf{H} \bm{\Delta}_{b'}^\mathsf{H} \bm{\Phi}_{b'}^\mathsf{H}     \in \mathbb{C}^{n_{\text{A}} \times N_{\text{B}}}$, with rank denoted by $r_{b'}$, $\forall b'\neq b$. More specifically, we collect all the reduced-dimension effective matrices of unintended users and stack them in a column-wise fashion as 
\begin{align}
    \bm{\Xi}_b &= \Big[\bm{\Phi}_{\text{A}}^\mathsf{H} \bH_{1}^\mathsf{H}, \cdots, \bm{\Phi}_{\text{A}}^\mathsf{H}\bH_{b-1}^\mathsf{H},\bm{\Phi}_{\text{A}}^\mathsf{H}\bH_{b+1}^\mathsf{H}, \cdots, \bm{\Phi}_{\text{A}}^\mathsf{H}\bH_{B}^\mathsf{H}\Big]
    %\in \mathbb{C}^{n_{\text{A}} \times \sum_{b'\neq b, b'\in \mathcal{B}} N_{\text{B}}}
    ,
\end{align}
for $b \in \mathcal{B}$, with a rank $\bar{r}_b = \sum\limits_{b'\in \mathcal{B}, b' \neq b} r_{b'}$. Then, given that $\bar{r}_b < B N_{\text{B}}, \forall b \in \mathcal{B}$ due to the correlated entries of $\bm{\Phi}_{\text{A}}^\mathsf{H}\bH_{b}^\mathsf{H}$, the beamformer $\bp_b \in \mathbb{C}^{ n_{\text{A}}}$ can be obtained from the orthonormal basis of the nontrivial null space of $\bm{\Xi}_b$, which we can choose from the left singular vectors of $\bm{\Xi}_b$ that are associated with zero 
singular values. To this end, we perform singular value decomposition (SVD) and write
\begin{align}
    \bm{\Xi}_b &= \begin{bmatrix} \bU^{(1)}_b & \bU^{(0)}_b \end{bmatrix} \begin{bmatrix} \bm{\Omega}^{(1)}_b & \mathbf{0} \\ \mathbf{0} & \bm{\Omega}^{(0)}_b \end{bmatrix}
      \bV^\mathsf{H}_b,
\end{align}
where $\bm{\Omega}^{(1)}_b$ and $\bm{\Omega}^{(0)}_b$ are diagonal matrices that comprise the nonzero and zero singular values of $\bm{\Xi}_b$, respectively, $\bU^{(1)}_b$ and $\bU^{(0)}_b$ are semi-unitary matrices that comprise the corresponding left singular vectors, and $\bV_b$ comprises the right singular vectors of $\bm{\Xi}_b$. More specifically, given that the matrix $\bU^{(0)}_b \in \mathbb{C}^{n_{\text{A}} \times (n_{\text{A}} - \bar{r}_b)}$ comprises $n_{\text{A}} - \bar{r}_b$ orthonormal basis vectors of the null space of $\bm{\Xi}_b$, the desired inner beamformer can be 
\begin{align}
    \bp_b = \left[ \bU^{(0)}_b \right]_{:,1} \in \mathbb{C}^{n_{\text{A}}},
\end{align}
which satisfies $\| \bp_b \|_2^2 = 1$ and $\bH_{b'} \bm{\Phi}_{\text{A}} \bp_b = \bm{0}, \forall b \neq b' \in \mathcal{B}$, as long as the rank $\bar{r}_b < B N_{\text{B}}$ and the constraints $n_{\text{A}} > \bar{r}_b$ and $n_{\text{A}} - \bar{r}_b \geq 1$ are met.

\subsection{Receive Filter Design}
With the beamformer design presented in the previous subsection, all inter-user interference among the Bobs can be eliminated. This fact allows the $b$-th Bob to exploit its effective channel $\bH_b \bf_{b} = \bm{\Phi}_{b} \bm{\Delta}_b  \bG_{b}  \bm{\Delta}_{\text{A}} \bp_b \in \mathbb{C}^{n_{\text{A}}}$ for computing its reception combining vector, as follows
\begin{align}\label{zf_rec_mtx}
    \bq_b = \frac{\bm{\Phi}_{b} \bm{\Delta}_b  \bG_{b}  \bm{\Delta}_{\text{A}} \bp_b }{\left\|\bm{\Phi}_{b} \bm{\Delta}_b  \bG_{b}  \bm{\Delta}_{\text{A}} \bp_b \right\|_2} \in \mathbb{C}^{ n_{\text{A}}},
\end{align}
which is a matched filter vector, satisfying $\| \bq_b \|_2^2 = 1$, constructed based on the effective channel matrix observed only by the $b$-th Bob. It has a reduced dimension that is determined by the number of receive antennas $N_{\text{B}}$, where we adopt $N_{\text{B}} \ll N_{\text{A}}$ in this work. This property makes the proposed approach much less demanding than relying on the full channel matrix $\bH_b \in \mathbb{C}^{N_{\text{B}} \times N_{\text{A}}}$, which has a much higher dimension.
By employing the reception combining vector in Eq.~\eqref{zf_rec_mtx}, the $b$-th legitimate user will have the post-processed signal as  
\begin{align}\label{rec_sig_bobs}
    y_b &= \underset{\text{Signal of interest}}{\underbrace{\bq_b^\mathsf{H}\bH_b\bf_b\sqrt{\zeta_b\alpha_{b}} s_b}} + \underset{\text{Artificial noise}}{\underbrace{\bq_b^\mathsf{H}\bH_b\bf_b \sqrt{\zeta_b\beta_{b}} w_b}} + \hspace{-0.4cm}\underset{\text{Additive noise}}{\underbrace{\bq_b^\mathsf{H}\bz_b}}.
\end{align}

On the other hand, we assume that Eve infiltrates into the system and gets access to the effective channels $\bH_{\text{E}}\bf_{b}, \forall b\in \mathcal{B}$ of the legitimate users. Note, however, that because $\bf_b, \forall b\in \mathcal{B}$, is computed based on the channels of legitimate users only, the signals intended for other Bobs, i.e., $\forall b^\star\in \mathcal{B}, b^\star \neq b$, will cause interference to Eve when Eve eavesdrops on the $b$-th Bob.
%, i.e., $b^\star = \arg \min_{\forall b \in \mathcal{B}} \|\bH_{\text{E}} - \bH_b\|_F$.
In addition, Eve is not aware that Alice is transmitting AN. Under these assumptions, Eve computes its reception vector $\bq_{\text{E}}$ following the same approach as in Eq.~\eqref{zf_rec_mtx} but based on Eve's effective channel matrix $\bH_{\text{E}} \bf_{b} \in \mathbb{C}^{n_{\text{A}}}$ associated with the target user $b$. More specifically, Eve's receive combining vector is obtained as
\begin{align}\label{zf_rec_mtx_eve}
    \bq_{\text{E}} = \frac{\bm{\Phi}_{\text{E}} \bm{\Delta}_{\text{E}}  \bG_{\text{E}}  \bm{\Delta}_{\text{A}} \bp_{b} }{\left\|\bm{\Phi}_{\text{E}} \bm{\Delta}_{\text{E}}  \bG_{\text{E}}  \bm{\Delta}_{\text{A}} \bp_{b} \right\|_2} \in \mathbb{C}^{ n_{\text{A}}}.
\end{align}
Then, after filtering the eavesdropped signal of user $b$ through $\bq_{\text{E}}$, Eve has the post-processed signal as
\begin{align}\label{rec_sig_eve}
    y_E &= \bq_{\text{E}}^\mathsf{H}\bH_{\text{E}} \sqrt{\zeta_{\text{E}}}\bigg(\hspace{1mm}\underset{\text{Signal of interest}}{\underbrace{\bf_{b}\sqrt{\alpha_{b}} s_{b}}} + \underset{\text{Artificial noise}}{\underbrace{\bf_{b} \sqrt{\beta_{b}} w_{b}}} \nonumber\\
    &+ \underset{\text{Inter-user interference}}{\underbrace{\sum_{b^\star\in \mathcal{B}, b^\star \neq b} \bf_{b^\star} (\sqrt{\alpha_{b^\star}} s_{b^\star} + \sqrt{\beta_{b^\star}} w_{b^\star})}} \hspace{1mm} \bigg) + \underset{\text{Additive noise}}{\underbrace{\bq_{\text{E}}^\mathsf{H}\bz_E}}. \nonumber
\end{align}

The corresponding signal-to-interference-plus-noise ratios (SINRs) as well as the secrecy capacity experienced in the system are investigated in the sequel.

\section{Secrecy Analysis and Power Allocation}

\subsection{SINR Expressions}
Alice informs all legitimate users of the exploitation of AN.  Therefore, we assume that $w_b$ can be successfully subtracted from the signal in Eq.~\eqref{rec_sig_bobs} with the aid of the successive interference cancellation (SIC) technique. As a result, the SINR observed by the $b$-th Bob when recovering its information symbol, for $\forall b \in \mathcal{B}$, can be given by
\begin{align}
    \gamma_b = \frac{|\bq_b^\mathsf{H}\bH_b\bf_b\sqrt{\zeta_b\alpha_b}|^2}{|\bq_b^\mathsf{H}\bz_b|^2} = \frac{|\bq_b^\mathsf{H}\bH_b\bf_b|^2 \zeta_b\alpha_b}{\sigma_z^2}.
\end{align}

In contrast to Bobs, Eve cannot decode the AN $w_b$ and, thus, it will be able to eavesdrop only on a noisy version of the transmitted information symbol, which is also corrupted by inter-user interference. To be specific, when detecting the symbol of the target user $b \in \mathcal{B}$, Eve observes the following SINR
\begin{align}\label{eq:gamma_b}
    \gamma^{b}_{\text{E}} = \frac{|\bq_{\text{E}}^\mathsf{H}\bH_{\text{E}}\bf_{b}|^2 \zeta_{\text{E}} \alpha_{b}}{\big(
    \underset{\scalebox{1}{$ + \sum_{b^\star\in \mathcal{B}, b^\star \neq b} |\bq_{\text{E}}^\mathsf{H}\bH_{\text{E}} \bf_{b^\star} |^2\zeta_{\text{E}}\beta_{b^\star} + \sigma_z^2$}\big)}{|\bq_{\text{E}}^\mathsf{H}\bH_{\text{E}}\bf_{b}|^2 \zeta_{\text{E}}\beta_{b} +  \sum_{b^\star\in \mathcal{B}, b^\star \neq b} | \bq_{\text{E}}^\mathsf{H}\bH_{\text{E}}\bf_{b^\star} |^2 \zeta_{\text{E}}\alpha_{b^\star}}},
\end{align}
where the numerator $|\bq_{\text{E}}^\mathsf{H}\bH_{\text{E}}\bf_{b}|^2 \zeta_{\text{E}} \alpha_{b}$ represents the received power of the signal of interest. The denominator, on the other hand, represents the total interference and noise power observed by Eve. It consists of four components: (i) the power of the AN intended for Bob $b$, $|\bq_{\text{E}}^\mathsf{H}\bH_{\text{E}}\bf_{b}|^2 \zeta_{\text{E}}\beta_{b}$, (ii) the sum powers of the signals intended for all the other legitimate users, $\sum_{b^\star\in \mathcal{B}, b^\star \neq b} | \bq_{\text{E}}^\mathsf{H}\bH_{\text{E}}\bf_{b^\star} |^2 \zeta_{\text{E}}\alpha_{b^\star}$, (iii) the sum powers of AN intended for all the other legitimate users,  $\sum_{b^\star\in \mathcal{B}, b^\star \neq b} |\bq_{\text{E}}^\mathsf{H}\bH_{\text{E}} \bf_{b^\star} |^2\zeta_{\text{E}}\beta_{b^\star}$, and (iv) the noise power $\sigma_z^2$.

\subsection{Secrecy Capacity}
With the above derivations of SINRs, the rates achieved by the $b$-th Bob and Eve are given by
$
R_{b} = \log_2\big(1 + \gamma_{b}\big),
$
and
$
    R^{b}_{\text{E}} = \log_2\big(1 + \gamma^{b}_{\text{E}}\big),
$
respectively. As a result, the secrecy capacity in bits per channel use (bpcu) observed for the legitimate user $b \in \mathcal{B}$ can be computed by
\begin{equation}
    S_{b} = \Big[ R_{b} - R^{b}_{\text{E}} \Big]^+,
\end{equation}
where $[a]^+ = \max\{a,0\}$.

\subsection{PA Formulation and Solution}
By following the MMF tradition, 
we aim to maximize the minimum of the secrecy rates of Bobs. The associated optimization problem can be formulated as follows:
\begin{subequations}\label{p1}
	\begin{align}
\mathcal{P}_1: \;\; &\underset{\alpha_{b},\beta_{b}}{\max} \hspace{2mm}\underset{\forall b \in \mathcal{B}}{\min} \left\{S_{b} \right\} \\
	&\text{s.t.}~
	 \sum_{b =1}^B \alpha_{b} + \beta_{b} = P_T,\label{p1b}\\
	& \alpha_{b} \geq 0, \beta_{b} \geq 0,\label{p1c}
	\end{align}
\end{subequations}
under the constraint of sum transmit power in~\eqref{p1b}. We conduct PA based on the instantaneous CSI, either perfect or imperfect. It is noted that the MMF form in the objective function makes the problem $\mathcal{P}_1$ intractable. To address this issue, we reformulate the optimization problem $\mathcal{P}_1$ as 
    \begin{subequations}  
	\begin{align}	
		& \mathcal{P}_2: \;\;\underset{\alpha_b,\beta_b}{\mathop{\max }}\, ~\tau \\
		&\text{s.t.}~
		\eqref{p1b},\eqref{p1c}, \nonumber \\
		& R_b-R_E^b \ge \tau,   \label{p2b}
	\end{align}
\end{subequations} 
by introducing the auxiliary variable $\tau$. However, the non-convexity of constraint~\eqref{p2b} remains an obstacle for solving problem $\mathcal{P}_2$. To address this issue, we introduce an auxiliary variable $C_E^b$ to transform~\eqref{p2b} into the following two constraints: i.e., $R_b-C_E^b \ge \tau$ and $C_E^b-R_E^b \ge 0 $. The latter is non-convex, and can be further transformed into $2^{C_E^b}-1\ge |\bm q_E^\mathsf{H}\bm H_E\bm f_b|^2\zeta_E \frac{\alpha_b}{I_E^b}$ and $|\bm q_E^\mathsf{H}\bm H_E\bm f_b|^2\zeta_E\beta_b+\sum\limits_{b^\star=1,b^\star\neq b}^B |\bm q_E^\mathsf{H}\bm H_E\bm f_{b^\star}|^2\zeta_E\alpha_{b^\star}+\sum\limits_{b^\star=1,b^\star\neq b}^B |\bm q_E^\mathsf{H}\bm H_E\bm f_{b^\star}|^2\zeta_E\beta_{b^\star}+\sigma_z^2\ge I_E^b$, where $I_E^b$ is another newly introduced auxiliary variable. The tight coupling of optimization variables in the constraint $2^{C_E^b}-1\ge |\bm q_E^\mathsf{H}\bm H_E\bm f_b|^2\zeta_E \frac{\alpha_b}{I_E^b}$ introduces further challenges
in solving the optimization problem $\mathcal{P}_2$. By introducing a series of auxiliary variables, i.e., $X_b$, $Y_b$, and $Z_b$, for $b \in \mathcal{B}$, we can further transform it into four constraints, i.e., $\exp(Z) \ge |\bm q_E^\mathsf{H}\bm H_E\bm f_b|^2\zeta_E {\exp(X_b-Y_b)}$, $\alpha_b \le \exp(X_b)$, $I_E^b \ge \exp(Y_b)$, and $2^{C_E^b}-1\ge \exp(Z_b)$. Summarizing the above steps, the optimization problem $\mathcal{P}2$ becomes 
\begin{subequations}  
	\begin{align}	
		& \mathcal{P}_3: \;\;\underset{\alpha_b,\beta_b, C_E^b, I_E^b, X_b, Y_b,Z_b}{\mathop{\max }}\, ~\tau \\
		&\text{s.t.}~
		\eqref{p1b},\eqref{p1c},\nonumber\\
		&R_b-C_E^b \ge \tau,\label{p3b}\\
   		&\exp(Z_b) \ge |\bm q_E^\mathsf{H}\bm H_E\bm f_b|^2\zeta_E {\exp(X_b-Y_b)}, \label{p3c}\\
       &{\sum\limits_{i=1,i\neq b}^B |\bm q_E^\mathsf{H}\bm H_E\bm f_i|^2\zeta_E\alpha_i \hspace{-0.1cm}+\hspace{-0.12cm}\sum\limits_{i=1}^B |\bm q_E^\mathsf{H}\bm H_E\bm f_i|^2\zeta_E\beta_i\hspace{-0.1cm}+\hspace{-0.10cm}\sigma_z^2}\ge I_E^b, \label{p3d}\\
        &\alpha_b \le \exp(X_b),\label{p3e}\\
       &I_E^b \ge \exp(Y_b),\label{p3f}\\
       &2^{C_E^b}-1\ge \exp(Z_b).\label{p3g}
	\end{align}
\end{subequations}  
Although the problem $\mathcal{P}_3$ becomes more tractable than the original problem $\mathcal{P}_1$, constraints
\eqref{p3e} and \eqref{p3g}  remain non-convex. To address this, we employ the successive convex approximation (SCA) method with
first-order Taylor expansion to tackle them. In particular, the problem $\mathcal{P}_3$ can be finally rewritten as
\begin{subequations}  
	\begin{align}	
		& \mathcal{P}_4: \;\;\underset{\alpha_b,\beta_b,  C_E^b, I_E^b, X_b, Y_b,Z_b }{\mathop{\max }}\, ~ \tau \\
		&\text{s.t.}~
		\eqref{p1b},\eqref{p1c}, \eqref{p3b}, \eqref{p3c},\eqref{p3d}, \eqref{p3f},\nonumber\\
		&\alpha_b \le \exp(\bar X_b[n])(X_b-\bar X_b[n]+1),\label{Eq:First_order_approx_1}\\
		&2^{\bar C_E^b[n]}(\ln 2(C_E^b-\bar C_E^b[n])+1)-1\ge \exp(Z_b). \label{Eq:First_order_approx_2}
	\end{align}
\end{subequations}  
The right-hand side of ~\eqref{Eq:First_order_approx_1} and the left-hand side of~\eqref{Eq:First_order_approx_2} are the first-order approximations of $\exp(X_b)$ and $2^{C_E^b}-1$ at points $\bar X_b[n]$ and $\bar C_E^b[n]$, respectively, which are the solutions of $X_b$ and $C_E^b$ from the $n$-th iteration. Obviously, $\mathcal{P}_4$ is a convex problem that can be solved by using the well-known Matlab CVX toolbox~\cite{CVX}.  
% \rev{I THINK I CAN OBTAIN THE SOLUTION FOR THE ABOVE PROBLEM IN CLOSED-FORM.}

% \rev{I AM NOT SURE, HOWEVER, IF THIS  OBJECTIVE IS ENOUGH, WE CAN MAKE IT MORE COMPLICATED, WE CAN DISCUSS THIS.}

% \clearpage
% \newpage

% \rev{$>>$ Consider as a draft from this point onward $<<$}

% \section{Secrecy Capacity Optimization}

\section{Simulation Results}\label{sec_Sim_results}

\begin{figure}[t]
	\centering
\includegraphics[width=.60\linewidth]{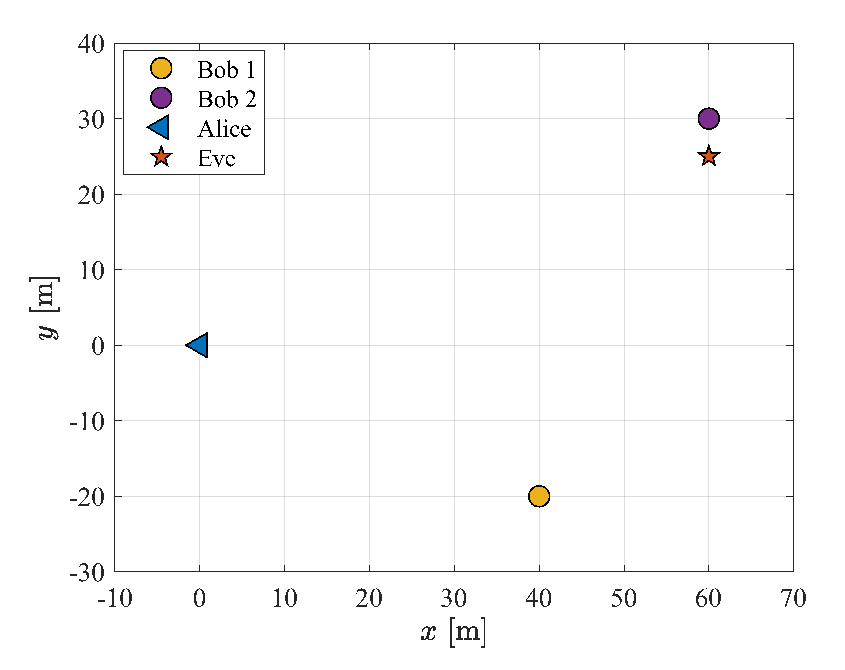}
	\caption{Simulation setup.}		\label{fig:Simulation_setup}
\end{figure}

\begin{figure}[t]
	\centering
\includegraphics[width=.8\linewidth]{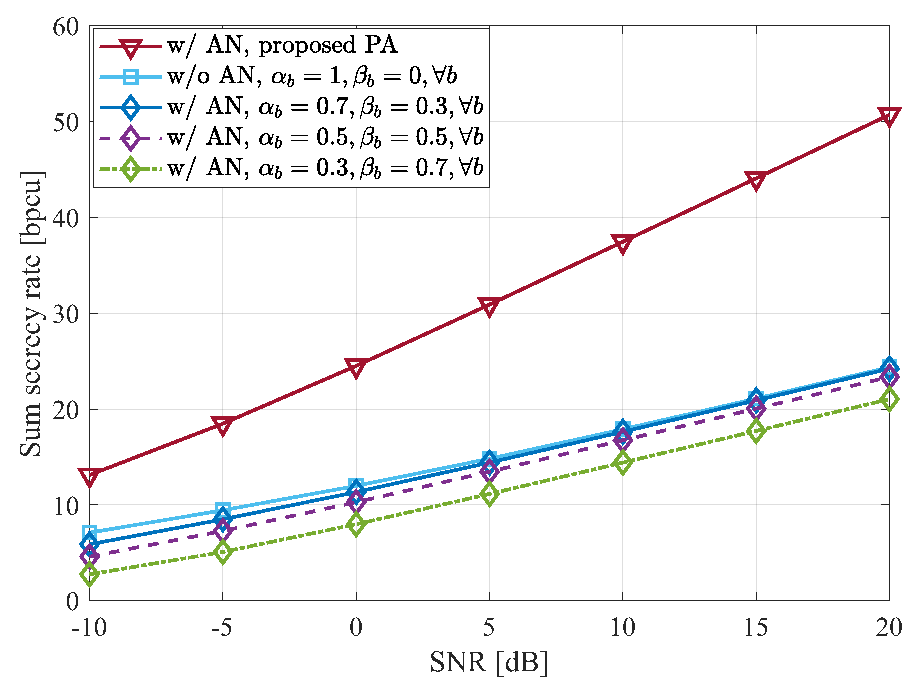}
	\caption{Sum secrecy capacity with the proposed PA ($N_{\text{B}} = N_{\text{E}} = 10\times 10$, channel error variance $\xi = 0$, and $\delta = \lambda/4$).}
		\label{fig:Effect_power_frac}
\end{figure}

% \begin{figure}[t]
% 	\centering
% \includegraphics[width=.8\linewidth]{res3.eps}
% 	\caption{Ergodic capacity for targeted user ($N_{\text{B}} = N_{\text{E}} = 10\times 10$, channel error variance $\xi = 0$)}
% 		\label{r3}
% \end{figure}

\begin{figure}[t]
	\centering
\includegraphics[width=.8\linewidth]{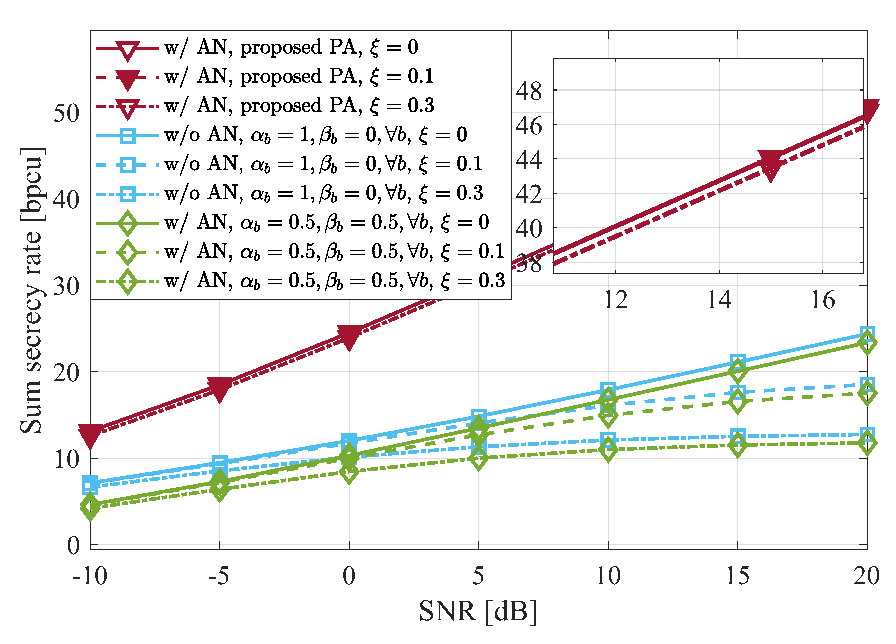}
	\caption{Effect of CSI imperfectness on sum secrecy capacity ($N_{\text{B}} = N_{\text{E}} = 10\times 10$, various channel error variances $\xi$'s).}
		\label{fig:Effect_channel_error}
\end{figure}

\begin{figure}[t]
	\centering
\includegraphics[width=.8\linewidth]{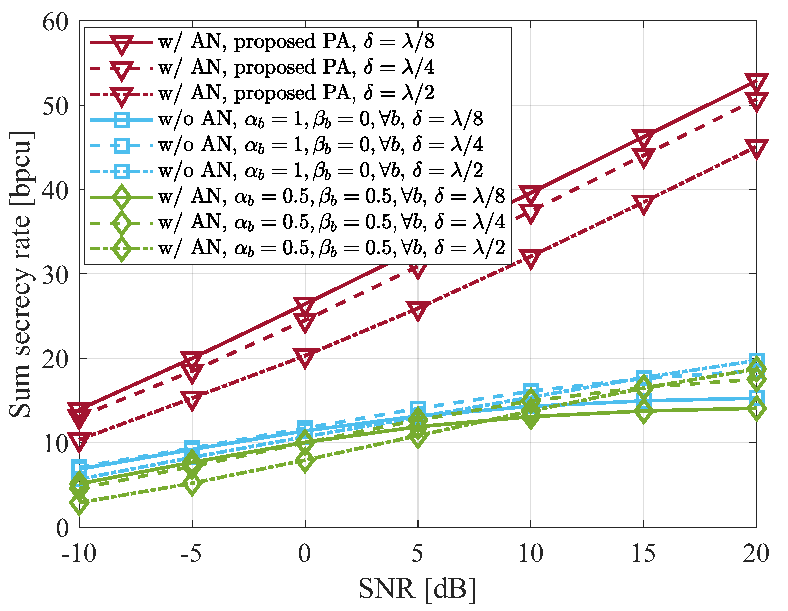}
	\caption{Effect of inter element spacing on sum secrecy capacity ($N_{\text{B}} = N_{\text{E}} 
 = 10\times 10$, various $\delta$'s, channel error variance $\xi = 0.1$).}
		\label{fig:Effect_inter_element_spacing}
\end{figure}

\begin{figure}[t]
	\centering
\includegraphics[width=.8\linewidth]{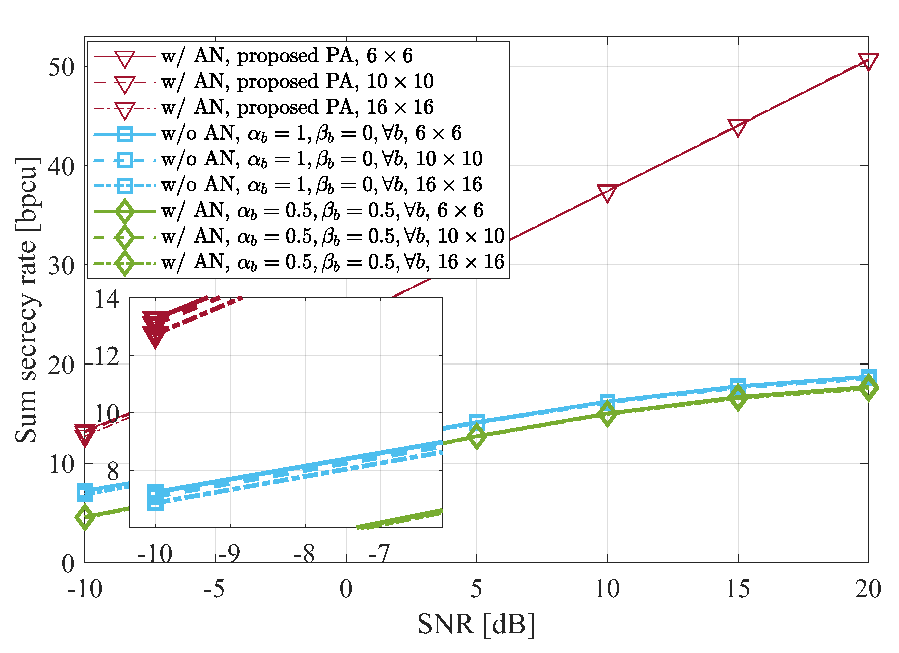}
	\caption{Effect of the number of Eve's antennas on sum secrecy capacity ($N_{\text{B}} = 10\times 10$, various $N_{\text{E}}$'s, channel error variance $\xi = 0.1$).}
		\label{fig:Effect_Eve_antennas}
\end{figure}

In this section, we present the results of the analysis and optimization of the sum secrecy rate of HMIMO with PA. The baseline scheme with fixed PA coefficients is introduced and compared. To this end, we implement a scenario in which $B=2$ Bobs and one Eve receive information from one Alice. Unless otherwise stated, Alice and the Bobs employ, respectively, $N_{\text{A}} = 20 \times 20$ and $N_{\text{B}} = 10 \times 10$ antenna elements. On the other hand, we test the effect of different numbers of antennas for Eve on the secrecy performance in the sequel. The first antenna element of Alice is located in the origin of the 3D plane, i.e., its coordinate being $\bp_{\text{A},1} = [0,0,0]$, whereas the 3D coordinates for the first antennas of Bobs $1$ and $2$ are $\bp_{1,1} = [40, -20,  0]$ and $\bp_{2,1} = [60, 30,  0]$, respectively. We assume that Eve is close to Bob $2$ with its first antenna located at $\bp_{E,1} = [60, 25,  0]$. Such a simulation setup is depicted in Fig.~\ref{fig:Simulation_setup}. For the channel parameters, we set $\eta = 2.7$ and $\Lambda = 1000$. Unless otherwise stated, the inter-element spacing of all arrays is set as $\delta = \lambda/4$, with antennas indexed by $\bp_{i,n} = [[\bp_{i,1}]_1 + \delta \cdot \mathrm{mod}(n-1, N_{i,x}), [\bp_{i,1}]_2 + \delta\cdot \lfloor (n-1)/N_{i,x} \rfloor, 0]$, for $n=2,\cdots, N_i$, with $i \in \{\text{A}, \mathcal{B}, \text{E}\}$. The sum transmit power is configured as $P_T = \sum_{b =1}^2 \alpha_b + \beta_b = 2$. The signal-to-noise ratio (SNR) is defined as $1/\sigma_z^2$ and the number of trials is set to $1000$.
% In these preliminary results, we provide examples considering only $n_S = 1$ symbol per user.

Fig.~\ref{fig:Effect_power_frac} compares the proposed PA scheme with fixed PA coefficients in terms of sum secrecy rate, showing the significant performance enhancement (up to two fold) with the aid of PA, especially in the high SNR regime. When the SNR value is $20$ dB, the proposed PA approach with $\delta = \lambda/8$ achieves more than $50$ bpcu while the fixed PA scheme with $\alpha_b = \beta_b = 0.5$ and $\delta = \lambda/8$ achieves about $24$ bpcu. For the fixed PA coefficients, the performance becomes better when $\alpha_b$ increases. However, this does not mean that AN fails to play a essential role in secrecy performance. With the aid of proposed PA, we are able to turn AN from foe to friend. Fig.~\ref{fig:Effect_channel_error} studies the effect of CSI imperfectness. For the case of fixed PA coefficients, the performance degradation is obvious when the SNR is large. However, the proposed PA scheme shows great robustness against the imperfectness of CSI. 
Fig.~\ref{fig:Effect_inter_element_spacing} examines the impact of inter-element spacing. It is noted that we keep the numbers of BS antennas unchanged. When the inter-element spacing reduces from $\lambda/2$ to $\lambda/8$, we observe performance gain in our proposed PA scheme. However, with fixed PA coefficients, only a small gain is observed in the low SNR regime, and it vanishes as the SNR increases. Fig.~\ref{fig:Effect_Eve_antennas} studies the effect of $N_\text{E}$ on the sum secrecy rate. Among the selected setups, e.g., $N_\text{E} \in \{6 \times 6, 10 \times 10, 16 \times 16\}$, the performance is almost overlapping. In other words, the number of Eve's antennas fails to play an essential role in the sum secrecy rate of the studied multi-user HMIMO systems. The reason lies in that the term related to Eve's combining vector $\bq_\text{E}$, i.e., $|\bq_{\text{E}}^\mathsf{H}\bH_{\text{E}}\bf_{b}|^2$, appears in both the denominator and the numerator of \eqref{eq:gamma_b}. In this sense, its effect on Eve's rate will be cancelled out, leaving the major impact from the control of $\alpha_b$ and $\beta_b$, i.e., power allocation.

\begin{figure*}[t]
	\centering
\includegraphics[width=1\linewidth]{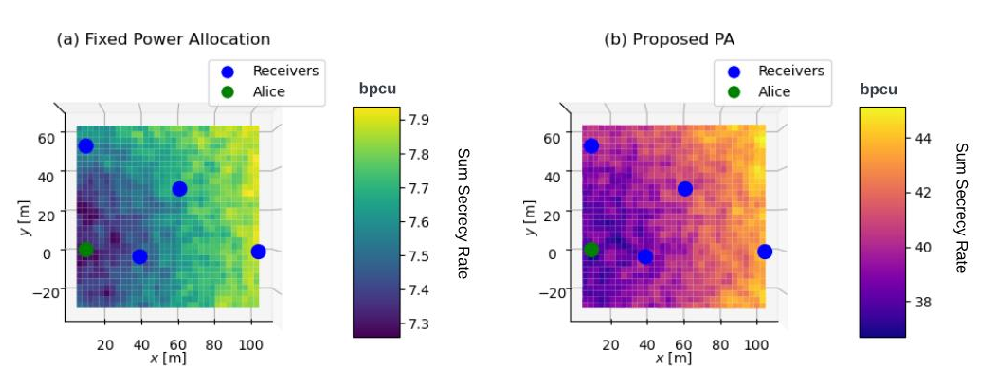}
	\caption{Heat map in terms of sum secrecy rate [bpcu] for four-Bob scenario: (a) fixed power allocation ($\forall b, \alpha_b = \beta_b = 0.5$), (b) proposed PA scheme, where the SNR value is set to be $-10$ dB.}
		\label{fig:Heat_map}
\end{figure*}

Last, we extend the two-Bob scenario to four-Bob scenario, where the four Bobs are spatially distributed over the $xy$ plane while fixing $z =0$. In this experiment, we introduce the heat-map to further illustrate the performance gain introduced by the proposed PA scheme compared to the fixed power allocation coefficient ($\forall b, \alpha_b = \beta_b = 0.5$) as a function of Eve's location (varying $x$ and $y$ coordinates while fixing $z =0$). The simulation results with the SNR being $-10$ dB are shown in Fig.~\ref{fig:Heat_map}. It is observed from the figure that the performance gain in terms of sum secrecy rate becomes more pronounced when the number of legitimate users increases (by comparing with Fig.~\ref{fig:Effect_power_frac}). This comes from the setup that the sum transmit power $P_T$ increases linearly as the number of legitimate users. The sum secrecy rate of the fixed PA scheme falls within the range $[7.25, 7.95]$ bpcu while that of the proposed PA falls within the range $[36, 46]$ bpcu. In addition, the proposed PA is insensitive to the location of Eve. In other words, regardless of the distance between Eve and one of the Bobs, the sum secrecy rates surrounding a specific legitimate user with the proposed PA are almost constant with a very small variation. 
% \begin{figure}[t]
% 	\centering
% \includegraphics[width=.91\linewidth]{Simulation_setup_four_bobs.eps}
% 	\caption{Simulation setup extended to four-Bob scenario.}
% 		\label{fig:Simulation_setup_four_bobs}
% \end{figure}

\section{Conclusion}
In this paper, we have studied the secrecy performance analysis of the multi-user HMIMO network under the max-min fairness, where AN is adopted. We have further addressed the PA problem and studied the effect of multiple system parameters on the sum secrecy rate. It has been demonstrated that with the aid of PA, up to two-fold sum secrecy rate can be achieved compared to the case with fixed PA coefficients in the two-Bob scenario. This becomes more profound when we further increase the number of legitimate users. The obtained heat maps have shown that the sum secrecy rate of the proposed PA scheme falls within $[36, 46]$ bpcu compared to $[7.25, 7.95]$ bpcu for the fixed PA scheme.

\balance

%  \appendix
% \section{Proofs} \label{Proofs}

\bibliographystyle{IEEEtran}
\bibliography{IEEEabrv,Ref}

\begin{thebibliography}{10}
\providecommand{\url}[1]{#1}
\csname url@samestyle\endcsname
\providecommand{\newblock}{\relax}
\providecommand{\bibinfo}[2]{#2}
\providecommand{\BIBentrySTDinterwordspacing}{\spaceskip=0pt\relax}
\providecommand{\BIBentryALTinterwordstretchfactor}{4}
\providecommand{\BIBentryALTinterwordspacing}{\spaceskip=\fontdimen2\font plus
\BIBentryALTinterwordstretchfactor\fontdimen3\font minus \fontdimen4\font\relax}
\providecommand{\BIBforeignlanguage}[2]{{%
\expandafter\ifx\csname l@#1\endcsname\relax
\typeout{** WARNING: IEEEtran.bst: No hyphenation pattern has been}%
\typeout{** loaded for the language `#1'. Using the pattern for}%
\typeout{** the default language instead.}%
\else
\language=\csname l@#1\endcsname
\fi
#2}}
\providecommand{\BIBdecl}{\relax}
\BIBdecl

\bibitem{Zhou2009}
X.~Zhou and M.~R. McKay, ``Physical layer security with artificial noise: Secrecy capacity and optimal power allocation,'' in \emph{proc. 3rd International Conference on Signal Processing and Communication Systems}, 2009, pp. 1--5.

\bibitem{Oggier2011}
F.~Oggier and B.~Hassibi, ``The secrecy capacity of the {MIMO} wiretap channel,'' \emph{{IEEE} Trans. Inf. Theory}, vol.~57, no.~8, pp. 4961--4972, 2011.

\bibitem{Hong2020}
S.~Hong, C.~Pan, H.~Ren, K.~Wang, and A.~Nallanathan, ``Artificial-noise-aided secure {MIMO} wireless communications via intelligent reflecting surface,'' \emph{{IEEE} Trans. Commun.}, vol.~68, no.~12, pp. 7851--7866, 2020.

\bibitem{Rafieifar2023}
A.~Rafieifar, H.~Ahmadinejad, S.~Mohammad~Razavizadeh, and J.~He, ``Secure beamforming in multi-user multi-{IRS} millimeter wave systems,'' \emph{{IEEE} Trans. Wireless Commun.}, pp. 1--1, 2023.

\bibitem{Huang2020}
C.~Huang, S.~Hu, G.~C. Alexandropoulos, A.~Zappone, C.~Yuen, R.~Zhang, M.~D. Renzo, and M.~Debbah, ``Holographic {MIMO} surfaces for {6G} wireless networks: Opportunities, challenges, and trends,'' \emph{{IEEE} Wireless Commun.}, vol.~27, no.~5, pp. 118--125, 2020.

\bibitem{Pizzo21}
A.~Pizzo, L.~Sanguinetti, and T.~L. Marzetta, ``{Fourier plane-wave series expansion for holographic {MIMO} communications},'' \emph{IEEE Trans. Wireless Commun.}, vol.~21, no.~9, pp. 6890--6905, 2022.

\bibitem{Ji23}
R.~Ji, S.~Chen, C.~Huang, J.~Yang, W.~E.~I. Sha, Z.~Zhang, C.~Yuen, and M.~Debbah, ``{Extra DoF of near-field holographic MIMO communications leveraging evanescent waves},'' \emph{IEEE Wireless Commun. Lett.}, pp. 1--1, 2023.

\bibitem{Pizzo20}
A.~Pizzo, T.~L. Marzetta, and L.~Sanguinetti, ``Spatially-stationary model for holographic {MIMO} small-scale fading,'' \emph{IEEE J. Sel. Areas Commun.}, vol.~38, no.~9, pp. 1964--1979, 2020.

\bibitem{Demir2022}
O.~T. Demir, E.~Bj\"ornson, and L.~Sanguinetti, ``Channel modeling and channel estimation for holographic massive {MIMO} with planar arrays,'' \emph{{IEEE} Wireless Commun. Lett.}, vol.~11, no.~5, pp. 997--1001, 2022.

\bibitem{Wei2022}
L.~Wei, C.~Huang, G.~C. Alexandropoulos, Z.~Yang, J.~Yang, W.~E.~I. Sha, Z.~Zhang, M.~Debbah, and C.~Yuen, ``Tri-polarized holographic {MIMO} surfaces for near-field communications: Channel modeling and precoding design,'' \emph{{IEEE} Trans. Wireless Commun.}, pp. 1--1, 2023.

\bibitem{Yang2017}
H.~Yang and T.~L. Marzetta, ``Massive {MIMO} with max-min power control in line-of-sight propagation environment,'' \emph{{IEEE} Trans. Commun.}, vol.~65, no.~11, pp. 4685--4693, 2017.

\bibitem{CVX}
M.~Grant and S.~Boyd, ``{CVX}: Matlab software for disciplined convex programming, version 2.1,'' \url{http://cvxr.com/cvx}, Mar. 2014.

\end{thebibliography}

\end{document}